\documentclass[referee,useAMS,usenatbib]{mn2e}

\usepackage{graphicx}
\font\twelvei = cmmi10 scaled\magstep1
       \font\teni = cmmi10 
\font\mbf = cmmib10 scaled\magstep1
       \font\mbfs = cmmib10 \font\mbfss = cmmib10 scaled 833
\font\msybf = cmbsy10 scaled\magstep1
       \font\msybfs = cmbsy10 \font\msybfss = cmbsy10 scaled 833
\def\lsim{\raise0.3ex\hbox{$<$}\kern-0.75em{\lower0.65ex\hbox{$\sim$}}}
\def\gsim{\raise0.3ex\hbox{$>$}\kern-0.75em{\lower0.65ex\hbox{$\sim$}}}

\title[Oscillating Shocks in an Accretion Flow Around Black Holes]
{Hydrodynamic Simulations of Oscillating Shock Waves in a Sub-Keplerian Accretion Flow
Around Black Holes}
\author[Kinsuk Giri, Sandip K. Chakrabarti, Madan M. Samanta, Dongsu Ryu]
{Kinsuk Giri$^{1}$, Sandip K. Chakrabarti$^{1,2}$, Madan M. Samanta$^{2}$, D. Ryu$^3$\\
$^{1}$S. N. Bose National Centre for Basic Sciences, Salt Lake,
              Kolkata 700098, India\\
$^{2}$Indian Centre for Space Physics, Chalantika 43, Garia Station Rd., 
	     Kolkata, 700084, India\\
$^3$ Dept. of Astron. Sp. Science, Chungnam Natl. Univ., Daejeon 305-764, South Korea} 
\begin{document}

\date{}


\maketitle

\label{firstpage}

\begin{abstract}
We study the accretion processes on a black hole by numerical simulation.
We use a grid based finite difference code for this purpose. We scan the parameter
space spanned by the specific energy and the angular momentum and compare the 
time-dependent solutions with those obtained from theoretical considerations.
We found several important results (a) The time dependent flow behaves close to a constant height
model flow in the pre-shock region and a flow with vertical equilibrium in the 
post-shock region. (c) The infall time scale in the post-shock region is 
several times higher than the free-fall time scale. (b) There are two 
discontinuities in the flow, one being just outside of the inner sonic point. 
Turbulence plays a major role in determining the locations of these discontinuities. 
(d) The two discontinuities oscillate with two different frequencies and behave 
as a coupled harmonic oscillator. A Fourier analysis of the variation of the 
outer shock location indicates higher power at the lower frequency and lower power at the 
higher frequency. The opposite is true when the analysis of the inner shock is made.
These behaviours will have implications in the spectral and timing properties of 
black hole candidates. 
\end{abstract}


\section{Introduction}

Understanding hydrodynamic behaviour of matter in the immediate vicinity of a black hole
is extremely important as the emitted radiation intensity from the flow depends on the 
density and temperature at each flow element at each moment of time. Thus the spectral
and temporal properties of emitted radiation is directly determined by the 
hydrodynamic variables. Early known simulations were carried out by 
Hawley, Smarr \& Wilson (1984ab). However, the code may have been executed with 
flow parameters not giving rise to the steady shocks and as a result, the shocks
were found to be propagating outward (but see, Hawley and Smarr, 1986 where solutions with shock
wave have been presented).  Chakrabarti \& Molteni (1993) and Molteni,
Lanzafame \& Chakrabarti (1994) used smoothed particle hydrodynamics (which preserves
the specific angular momentum accurately because the `particles' are toroidal in shape) to simulate one and two dimensional 
sub-Keplerian flows and showed that flows become transonic and also produce standing shocks
where they are predicted by well established theory of transonic flows 
(Chakrabarti, 1990; hereafter C90). Formation of steady shocks were seen in simulations 
which uses a grid based (Total Variation Diminishing, or TVD) scheme also
(Molteni, Ryu \& Chakrabarti, 1996).  Of course, if the
initial condition of the flow is chosen to be random, the flow may show time dependent
behaviour as the steady transonic solution is possible only when the initial parameters
are from a definite region of the parameter space. This was shown to be the case
by Ryu et al. (1995) and Ryu, Chakrabarti \& Molteni (1997). Using the TVD code, the latter 
authors showed that if the Rankine-Hugoniot condition is not satisfied, the shock is likely to
oscillate. The oscillating shocks were also observed in presence of cooling
(Molteni, Sponholz, \& Chakrabarti,
1996; Chakrabarti, Acharyya \& Molteni, 2004) and were widely assumed to be 
the cause of the quasi-periodic oscillations (QPOs) observed in radiations emitted by 
accretion flows around black holes. Since the generic physical processes 
which cause the oscillations of the shocks, and therefore, the oscillations in the
emitted radiation are the same in both the stellar and the massive/super-massive
black holes, a thorough study of the nature of shock oscillations and the dependence
of the oscillation frequencies on flow parameters is essential.

In this paper, we present the results of a series of simulations which 
basically sample the entire region of the parameter space spanned by the 
specific energy and specific angular momentum (Chakrabarti, 1989; here after C89; 
C90), i.e., the parameter space relevant for non-dissipative, non-magnetic and axisymmetric
hydrodynamic accretion flows. We use an axisymmetric grid based TVD code for this purpose. We 
obtain a large number of very important results, which, to our knowledge, have never been
published before: (a) Though as per injection condition at the outer boundary, the 
pre-shock flow properties match with the theoretical results of a constant height 
inflow, to our surprise, the post-shock flow properties match closely with the theoretical
results obtained assuming a hybrid-model inflow (Chakrabarti, 1989). What this means is that shock 
locations from simulations may be somewhat different from those of the theoretical result
which assumes either the constant height or vertical equilibrium condition in the both sides of the
shock. (b) The average infall time scale from the post-shock region appears to be a 
few times longer than the free fall time scale due to turbulence.
(c) For majority of the flow parameters, the shocks are found to be stable,
though they may oscillate around a mean location.
(d) Because of the strong turbulence close to a black hole which is formed due to 
the interaction of the incoming wave and the flow bounced back from the centrifugal barrier,
a weak shock is formed closer to the black hole, though, both the normal outer shock and the inner shock
seem to oscillate with the same frequencies -- the outer shock shows more power at lower
oscillation  frequency and the inner shock shows more power at higher oscillation frequencies.

The plan of our paper is the following:
In the next Section, we present the procedure of the total variation diminishing (TVD) method.
This is similar to what is used in Ryu et al. (1997) paper. In \S 3, we test the code with a simple 
zero angular momentum solution. In Molteni, Ryu \& Chakrabarti (1996), some test 
results with shock solutions were  already 
presented. Present we discuss results of a large number of simulations 
and compare the results.  Finally in \S 4, we make concluding remarks.

\section{Methodology of the Numerical Simulations}

The setup of our simulation is the same as in Ryu, Chakrabarti \& Molteni (1997).
We consider the adiabatic hydrodynamics of axisymmetric flows of gas under the 
Newtonian gravitational field of a point mass $M_{bh}$
located at the centre in cylindrical coordinates  $[R,\theta, Z]$. 
We also assume that the gravitational field of the black hole can be described 
by Paczy\'{n}ski \& Wiita (1980),
$$
\phi(R,Z) = -{GM_{bh}\over r-r_g} ,
\eqno{(1)}
$$
where, $r=\sqrt{R^2+Z^2}$ and the Schwarzschild radius is given by $r_g=2GM_{bh}/c^2$.
The black hole mass $M_{bh}$, the speed of light $c$, and the Schwarzschild radius $r_g$ 
are assumed to be the units of the mass, the velocity, and the length respectively. We use 
$x=R/r_g$, and $z=Z/r_g$ as the dimensionless  distances along x-axis and z-axis in the rest of the paper.
We also assume a polytropic equation of state for the
accreting (or, outflowing) matter, $P=K \rho^{\gamma}$, where,
$P$ and $\rho$ are the isotropic pressure and the matter density
respectively, $\gamma$ is the adiabatic index (assumed in this
paper to be constant throughout the flow, and is related to the
polytropic index $n$ by $\gamma = 1 + 1/n$) and $K$ is related
to the specific entropy of the flow. Since we ignore dissipation, the specific 
angular momentum $\lambda\equiv xv_{\theta}$ is  constant everywhere.
In C89 and C90, the steady solutions were classified according to the 
conserved flow parameters: the specific energy ${\cal E}$ and the specific 
angular momentum $\lambda$. At the outer boundary, where matter in injected, 
the velocity $v_x$ is supplied and the sound speed $a$ is computed from ${\cal E}$ from
the dimensionless conserved energy:
$$
{\cal E} = \frac{v_x^2}{2}+\frac{a^2}{\gamma-1}+\frac{\lambda ^2}{2x^2}+g(x).
\eqno{(2)}
$$
From Eq. (2), $g(x) = - 1/2(x-1)^{-1}$. We assume a fixed Mach number $M=v_x/a=10$  at the 
outer boundary and $z_o/x_o =0.1$, where, $z_o$ and $x_o$ are the height and the radial distance 
of the injected matter at the outer boundary.

The numerical simulation code is grid based using the TVD scheme, originally developed by Harten (1983).
It is an explicit, second order accurate scheme which
is designed to solve a hyperbolic system of the conservation equations,
like the system of the hydrodynamic conservation equations.
It is a nonlinear scheme obtained by first modifying the flux function
and then applying a non-oscillatory first order accurate scheme to
get a resulting second order accuracy. In this way we achieve the high resolution of a
second order accuracy while preserving the robustness of a non-oscillatory
first order scheme. The details of the code development are already in
Ryu et al. (1995, 1997) and are not presented here.

The calculations have been done in a setting similar to that used in 
Ryu, Chakrabarti \& Molteni (1997).  The computational box occupies one quadrant of the $x-z$ plane
with $0\leq x\leq50$ and $0\leq z\leq50$. 
We mimic the horizon ($x=1$) by placing an absorbing boundary at a sphere of radius $1.5$
inside which all the material is completely absorbed.
To begin the simulation, we fill in the black hole surroundings
with a very tenuous plasma of density $\rho_{bg}=10^{-6}$ and the temperature as
that of the incoming material. The incoming gas of density $\rho_{in}=1$
enters the box through the outer boundary located at
$x_o=50$. The adiabatic index $\gamma=4/3$ is chosen.
In the absence of self-gravity and cooling, the density is scaled out,
and thus the simulation results remain valid for accretion rate.
At the outer boundary, $x_o=50$ and thus $z_o = 5$. All the calculations 
have been done with $256\times256$ cells. Thus, each grid size is $\Delta x= \Delta z= 0.195$.

All the simulations are carried out assuming a stellar mass black hole ($M=10M_\odot$).
The results remain equally valid for massive/super-massive black holes, only the
time and the length are to be scaled with the central mass. We carry out the 
simulations till several thousands of dynamical time scales is passed. In
reality this corresponds to a few seconds in physical units. 

\section{Simulation Results}

In Molteni et al. (1996), we presented the simulation results where the steady state  
solutions from the theoretical and the numerical methods were compared. Generally, the 
solutions were reproduced quite well and we do not repeat the tests any more.

First, we run a model where the injected matter has no specific angular momentum $\lambda=0$. 
and follow the evolution. We compare the numerical solution with three 
models which are respectively the flow with a constant height (CH), the flow with an wedge 
shaped cross-section (constant angle: CA) and the flow which is in vertical
equilibrium (VE). See, Chakrabarti \& Das (2001) for the definitions of these models.
Here, the specific energy is chosen to be ${\cal E} = 0.023$. 
The flow approaches the black hole very smoothly and supersonically. We find that the numerical 
solution on the equatorial plane agrees very well with the theoretical results 
obtained with a constant height flow but not with the other two model solutions.

\begin{figure}
\includegraphics[height=12truecm,width=12truecm]{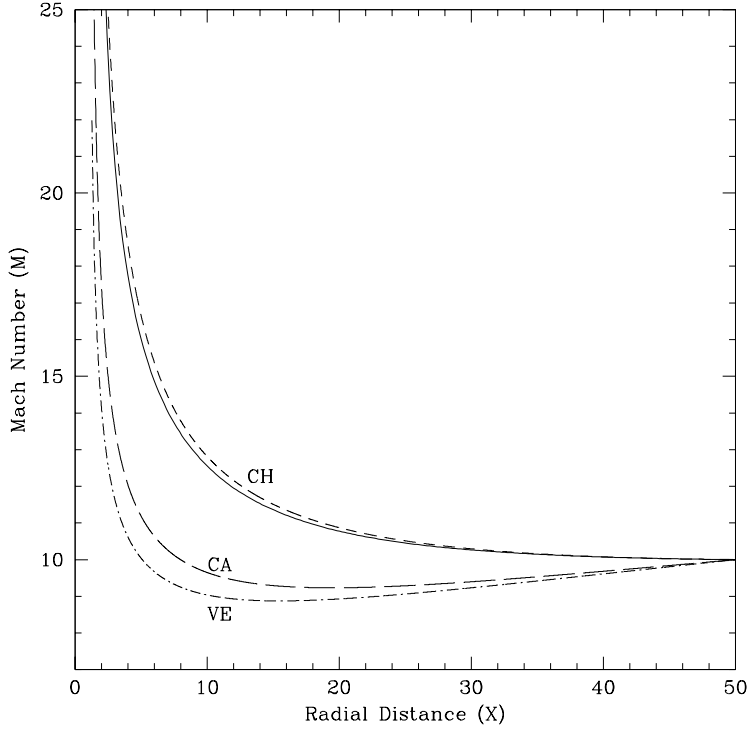}
\caption{The variation of the Mach number of a $\lambda = 0.0$, 
$e = 0.023$ flow as a function of the radial distance from the black hole. 
The solid curve is the solution obtained from the simulation, while the 
other curves are the theoretical results for constant height (CH), constant angle (CA)
and vertical equilibrium (VE) model. The time dependent solution agrees well with that 
of a constant height model flow.}
\end{figure} 

In the next set of simulations,  we include angular momentum. 
In Fig. 2, we present the classification of the
solutions in the parameter space spanned by the conserved energy and angular momentum (C89, C90).
In each region, the solution is qualitatively different. The classifications are made
in all the three models, CH, CA and VE. For the detailed meaning of various 
divisions in the parameter space, see, Chakrabarti (1996). Briefly, for CH model, the curve
$abc$ denotes the boundary between one (saddle type, on the left of the curve)
and two sonic points (one saddle type and one circle type) in the flow solution. 
The region $abd$ contains
flow parameters which produce two saddle type and one circle type sonic points, but 
no steady shock condition is satisfied. The region $dae$ has the same flow topologies as in $abd$
but the steady shocks can form in accretion flows. The flow with parameters from $eaf$
can form steady shocks only in winds and outflows. The solution topology in the region
$fag$ is same as that in $eaf$, but the steady shock condition is not satisfied. The points above 
$ag$ has only one saddle type sonic point. The solutions from points 
in $bcedb$ have one saddle type and one 
circle type sonic points, however, the solutions do not extend to infinity. 
Flows with parameters from other regions do not have any kind of steady solution. Similar curves for
the other two models, namely, CA and VE have similar meanings.
We have shaded one region which produces standing shocks in each model.
The Cases (A-H) which have been run are given in Table 1 where the values of the conserved
energy and angular momentum (${\cal E}, \lambda$) are shown. These values are also marked 
inside Fig. 2 with filled circles to show that depending on the theoretical model, the same pair of flow parameters
may or may not produce standing shocks. Our goal is to find out which model
is vindicated by the numerical simulation results. In Table 1, we also present the 
locations of the inner and outer sonic points, if any, and the stable shock locations, if any.

\begin{figure}
\includegraphics[height=12truecm,width=12truecm,angle=0]{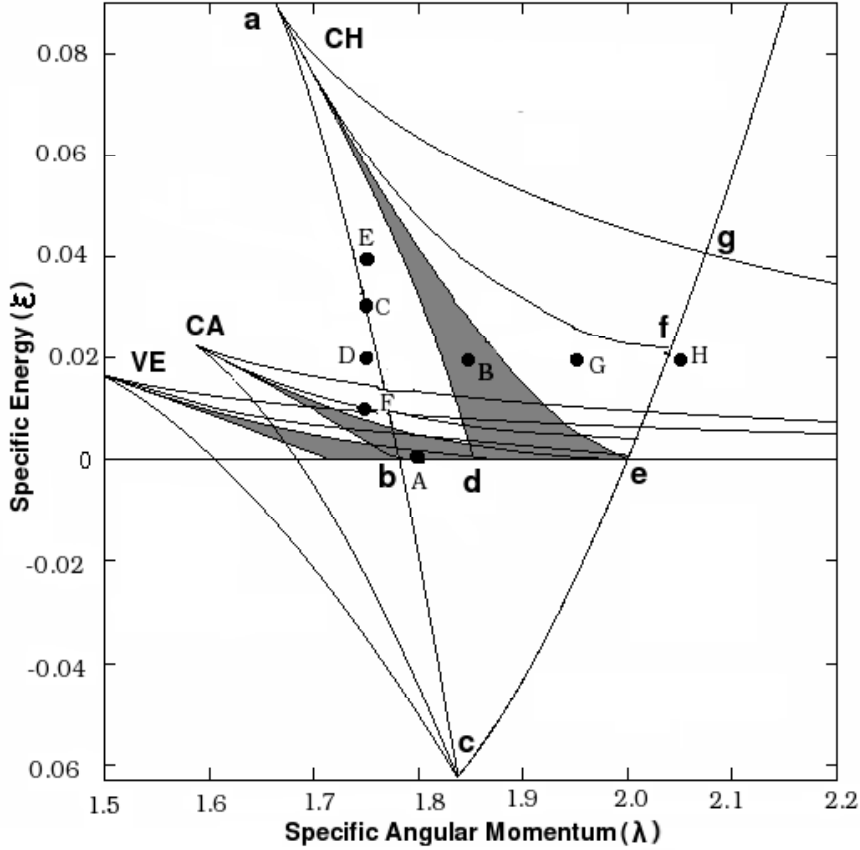}
\caption{The parameter pairs (${\cal E}, \lambda$) marked by filled circles
on the classification diagram are used for numerical 
simulations (Cases A-H) in this paper. See text for details.}
\end{figure} 

{\small
\begin{table}
\centering
\centerline {Table 1}
\centerline {Sonic points and shock locations (if any) obtained from three models for all the 
Model runs presented in this paper.}
\vskip 0.2cm

\begin{tabular}{|c|c|c|c|c|c|c|c|c|c|c|c|c|}

\hline \multicolumn{3}{|c|}{} & \multicolumn{3}{|c|}{Vertical Equilibrium} & \multicolumn{3}{|c|}{Wedge Shaped} &
\multicolumn{3}{|c|}{Constant Height} \\
\hline Case & ${\cal E}$ & $\lambda$ & $x_{in}$ & $x_{out}$ & $x_{s}$ & $x_{in}$ & $x_{out}$ & $x_{s}$ & $x_{in}$ & $x_{out}$ & $x_{s}$ \\ 
\hline A & 2.962e-04 & 1.80 & 2.321 & 1003.44 & 31.027 & 2.45 & 12.59 & 11.104 & 2.75 & 42.12 & $\relbar$ \\ 
\hline B & 0.02 & 1.85 & 2.154 & $\relbar$ & $\relbar$ & 2.24 & $\relbar$ & $\relbar$ & 2.36 & 56.1 & 11.42 \\ 
\hline C & 0.03 & 1.75 & 2.296 & $\relbar$ & $\relbar$ & 2.482 & $\relbar$ & $\relbar$ & 3.37 & 36.1 & $\relbar$ \\ 
\hline D & 0.02 & 1.75 & 2.341 & $\relbar$ & $\relbar$ & 2.53 & $\relbar$ & $\relbar$ & $\relbar$ & 57.162 & $\relbar$ \\ 
\hline E & 0.04 & 1.75 & 2.257 & $\relbar$ & $\relbar$ & 2.44 & $\relbar$ & $\relbar$ & 3.03 & 25.34 & $\relbar$ \\ 
\hline F & 0.01 & 1.75 & 2.393 & $\relbar$ & $\relbar$ & 2.60 & 28.825 & $\relbar$ & $\relbar$ & 119.87 & $\relbar$ \\ 
\hline G & 0.02 & 1.95 & 2.021 & $\relbar$ & $\relbar$ & 2.05 & $\relbar$ & $\relbar$ & 2.10 & $\relbar$ & $\relbar$ \\ 
\hline H & 0.02 & 2.05 & $\relbar$ & $\relbar$ & $\relbar$ & 2.06 & $\relbar$ & $\relbar$ & $\relbar$ & $\relbar$ & $\relbar$ \\ 
\hline 
\end{tabular} 
\end{table}
}

In Fig. 3, we show the results of Case A. The parameters used are $\lambda = 1.80$, ${\cal E} = 2.962 \times 10^{-4}$. 
The dotted curves give the variation of the Mach Number obtained from the numerical 
simulation (for the grid on the equatorial plane) at four consecutive times $1.90$s, 
$1.91$s, $1.92$s and $1.93$s. They are marked as 1, 2, 3 and 4 respectively.
Superposed on these curves are the theoretically obtained solutions for various models 
(marked) with the same outer boundary condition -- solid curves are for supersonic branch
and the long-dash-dotted curves are for subsonic branch. Theoretically, the steady shock is supposed to 
form at $31.027$ in vertical equilibrium model (Table 1). Numerically, however, we find that the flow has formed a
shock, but it is oscillating around a mean location. The flow Mach Number jumps and becomes subsonic at around ($x \sim 28$). 
However, the shock location oscillates. In some part in the post-shock region, the flow  has a 
`negative' Mach number. In this case, the matter actually flows outward, bouncing back from the 
centrifugal barrier on the equatorial plane. At around ($x \sim 7$) 
the flow becomes supersonic and again form a relatively weaker `inner shock' at around ($x \sim 5$). 
This inner shock also oscillates. 

Several important facts arise out of this exercise: (a) The inject matter behaves like a 
flow of constant thickness in the pre-shock region, (b) The Mach number variation in the
post-shock region is closer to that of the flow in vertical equilibrium. Of course, the 
back-flow due to the centrifugal barrier is a major factor to influence the post-shock 
region. This is clearly seen in Fig. 4 where the velocity vectors have been plotted. 
The back-flow diverts the matter from the post-shock region to regions away from the equatorial 
plane and produces jets and outflows. Some of these diverted matter
enters into the black hole from a height and becomes supersonic at around $2.5$.

\begin{figure}
\includegraphics[height=12truecm,width=12truecm]{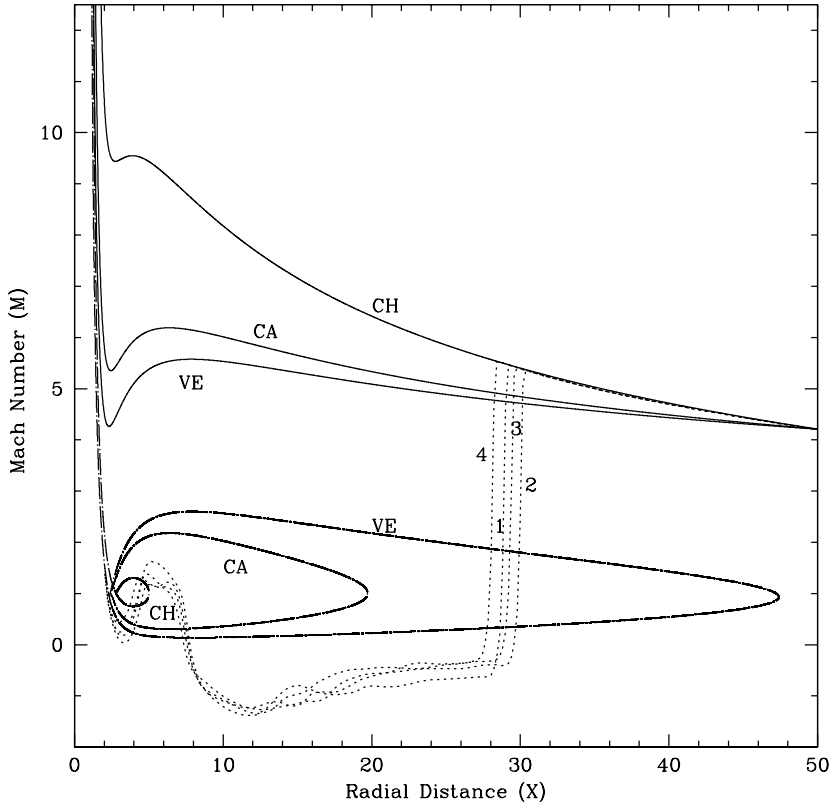}
\caption{{\bf A comparison of the theoretical results obtained from various models (marked)
with those from numerical simulations (dotted) for the same outer boundary condition (Case A). 
Solid curves are for the branch passing through outer sonic point and long dashed curves are for
the branch passing through the inner sonic point. Mach number variation on the equatorial 
plane is shown. The pre-shock region matches with that of a constant height flow, while the 
post-shock region is similar to the flow in vertical equilibrium. The presence of two shocks 
in the numerical solution may be noted.}}
\end{figure} 

\begin{figure}
\includegraphics[height=12truecm,width=12truecm,angle=270]{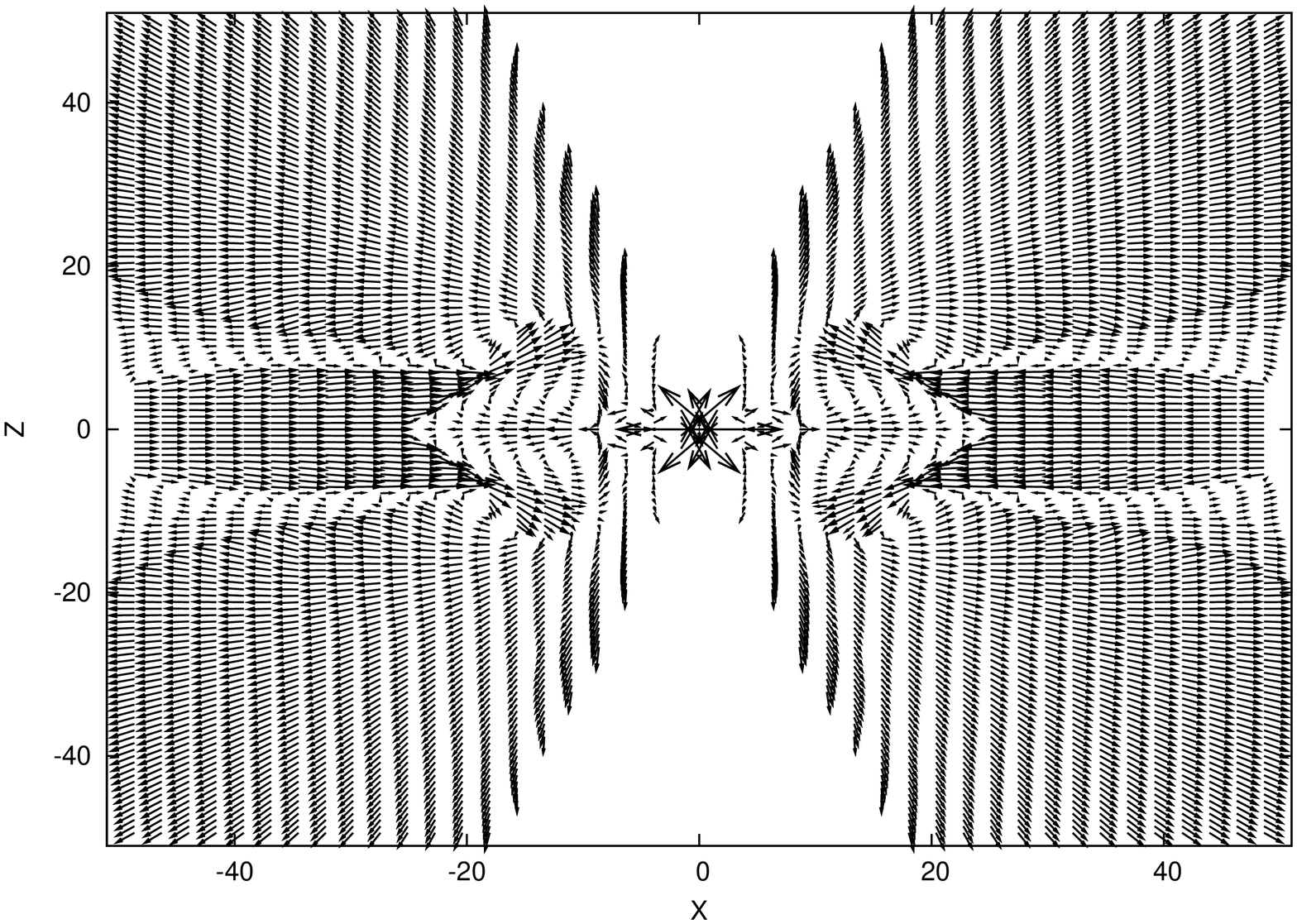}
\caption{The velocity vectors of matter obtained from the simulation for the Case A.
Due to centrifugal barrier, matter bounces backward and forms the shock. The 
injected flow in the post-shock region is deflected away from the 
equatorial region and enters into the black hole 
supersonically from higher elevation.}
\end{figure} 

The behaviour of the time dependent solution is evident in Figs. 5(a-d) where we
plot the velocity vector field and the density contours at regular intervals
at times $t=1.50$, $t=2.0$, $t=2.5$ and $3.0$s. The density contours in the 
post-shock region resemble those of a thick accretion disk (Paczynski \& Wiita, 1980),
though our results are more realistic since the radial velocity is included here.
The shock clearly moves around and the outflow also shows corresponding 
fluctuations. The density contours are for $0.01(0.01)0.1(0.1)1(1)13$, where
the density inside a parenthesis gives the interval and left and right numbers are 
`from' and `to' density values. The lowest density contour is at the highest altitude.
The presence of two oscillating shocks is clear from successive Figures. Some 
matter could be seen deflected outwards as outflows. The region close to the Z-axis 
remains empty due to the centrifugal barrier. Turbulence
slows down the flow and consequently, the infall time $t_{inf} =  \Sigma\frac{\delta r}{{v_r}}$,
where $\delta r$ is the grid size and $v_r$ is the local radial velocity,
is longer than the free-fall time $t_{ff}=r^{3/2}$ in the post-shock region. 
A detailed computation using the radial velocity averaged over $20$ grids
in the vertical direction and using radial coordinate from the outer shock to the
event horizon) shows that the ratio $R_t=t_{inf}/t_{ff}|_{post-shock} \sim 3.6$ in this case.
In the pre-shock region $R_t\sim 1$.

\begin{figure}
\includegraphics[height=12truecm,width=12truecm]{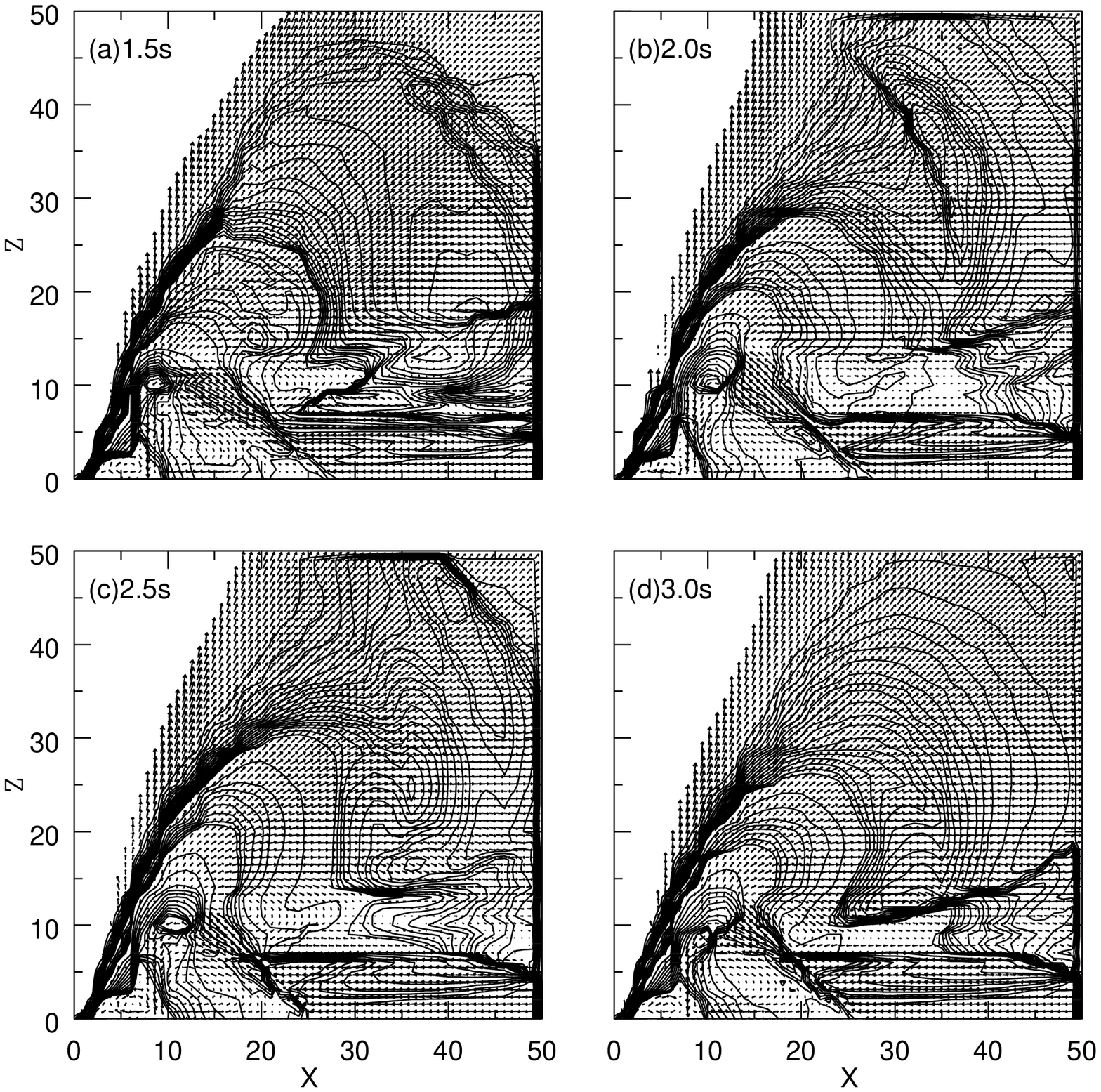}
\caption{Time dependent behaviour is best seen in the density contours and velocity vector fields plotted at a 
regular time intervals. Time in seconds is marked on each box. The flow is deflected at the shock contributing 
to jets and outflows. Both the shock locations could be seen.}
\end{figure} 

We continue our detailed investigation of the Case A. In Fig. 6, we show the variation of the outer (OS) 
and the inner shock (IS) locations (dimensionless unit) with time (in seconds). The oscillating nature 
settles down after an initial transient phase of $t \sim 0.06$. We clearly see the presence of 
oscillations in both the shocks, though the amplitudes are larger for the outer shock. The outer 
shock location oscillates between $26$ to $32$ and the inner shock oscillates between $3$ to $5$.   

In Fig. 7 (a-b), we present the power density spectrum (PDS) of the time variation of the shock 
locations. In (a) and (b), the PDSs for the inner and the outer shock locations are shown. The outer 
shock shows a peak at $1.56$Hz, but otherwise, both the PDSs show strong peaks at $\sim 18$Hz.
In the sub-sonic flow of the post-shock region, the movement of the inner shock also perturbs 
the outer shock location and thus the high frequencies are the same. As will be discussed later, these 
oscillations can cause significant modulations of the X-ray intensity and cause the so-called QPOs in 
black hole candidates (Molteni, Sponholz \& Chakrabarti, 1996; Chakrabarti, Acharyya \& Molteni, 2004).

\begin{figure}
\includegraphics[height=12truecm,width=12truecm]{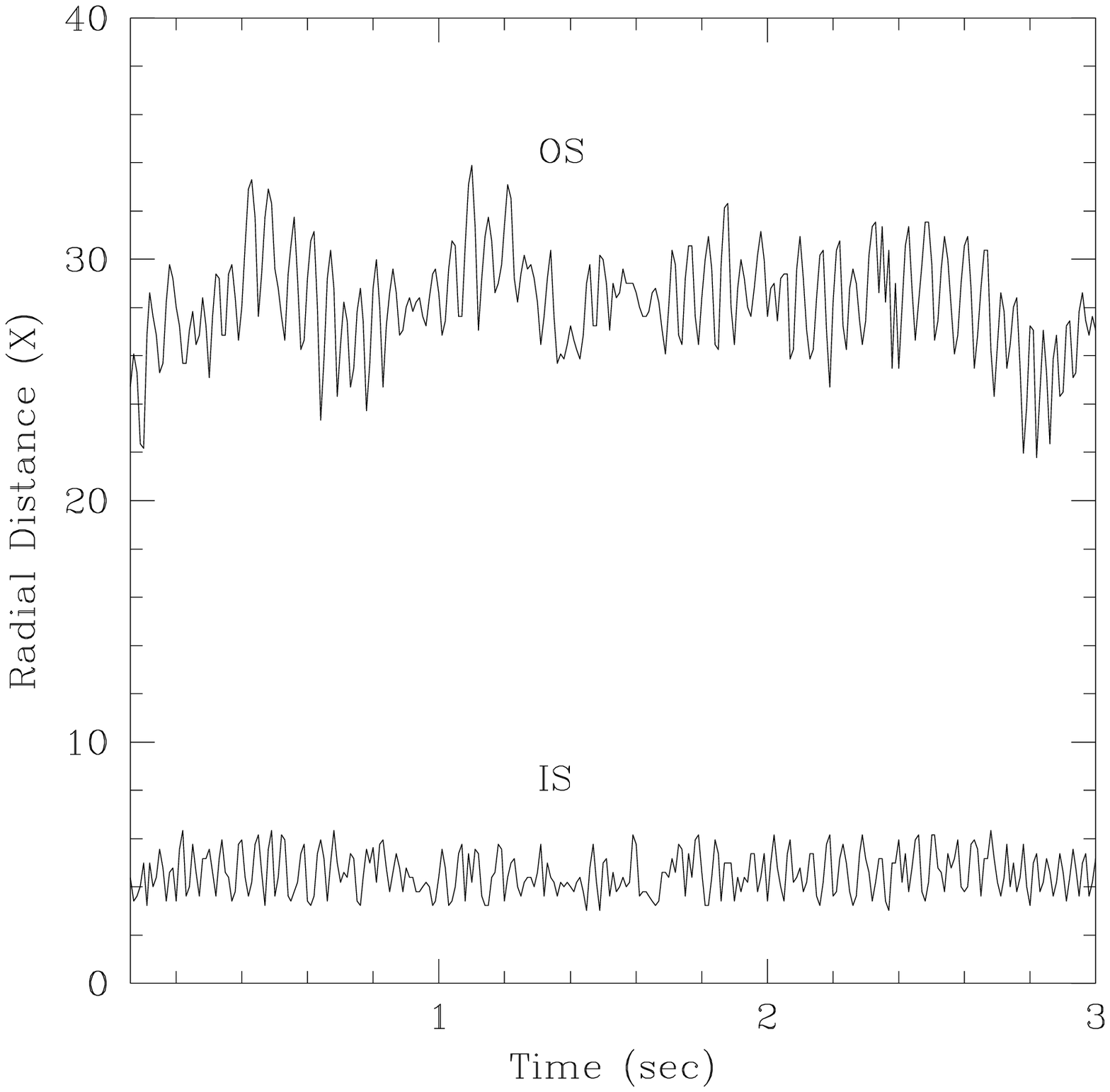}
\caption{Variation of the shock locations in dimensionless units with time (seconds). Case A parameters
are chosen for the simulation. A stellar mass $M=10M_\odot$ black hole was chosen for the purpose
of time computation. The time will scale with the mass of the black hole in this case since the
dissipation of the flow and the radiative transfer were neglected.}
\end{figure} 

\begin{figure}
\includegraphics[height=12truecm,width=12truecm,angle=0]{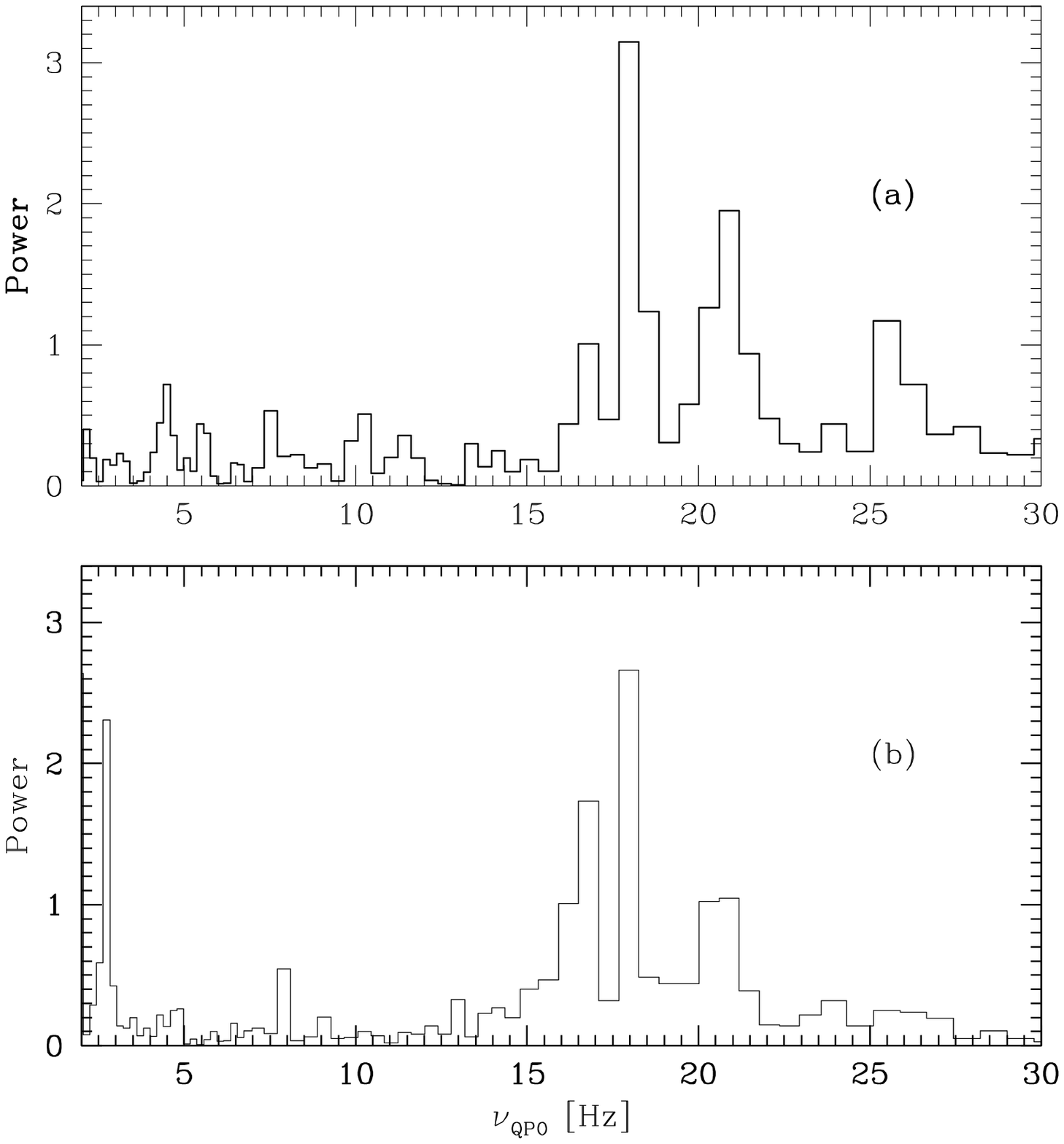}
\caption{{\bf(a-b)}: Power density spectra of the time variation of the (a) inner and the (b) outer shock
locations. The frequencies are calculated assuming $M=10M_\odot$. It will scale inversely
with the mass of the black hole. The outer shock shows strong peaks at $1.56$Hz and $17.97$Hz,
while the inner shock peaks at $17.97$Hz only. Such an oscillation is thought to cause the periodic 
modulations in X-ray intensities from black hole candidates.}
\end{figure} 

We now focus our attention to the oscillation properties of the shock locations by showing its dependence on 
the flow parameters. We choose the flow parameters of Cases D, B, G \& H (Table 1 and Fig. 2). We only 
vary $\lambda$ ($1.75, \ 1.85, \ 1.95$ and $2.05$ respectively) while keeping the specific energy the same. 
In Figs. 8(a-d), we compare the oscillations in the outer shock location and in Figs. 9(a-d), 
we compare the oscillations in the inner shock location. The mean outer shock location slightly 
increases when $\lambda$ increased from $1.75$ to $1.85$ but subsequently the shock becomes turbulent 
pressure dominated and not centrifugal pressure dominated and the location does not change even with angular 
momentum. The mean inner shock location, on the contrary, shows a tendency to increase with angular momentum. 
The PDS gives the frequencies of oscillations to be $\nu_{QPO}=2.54 (24.70),\ 2.93 (25.50),\ -- (28.61),\ -- (27.1)$ 
Hz for $\lambda=1.75, \ 1.85, \ 1.95$ and $2.05$ respectively for the outer (inner) shocks. The ratio $R_t$ of 
the infall time and the free fall time from the outer shock increases almost monotonically, which are $2.06$, $3.06$, 
$4.63$ and $4.52$ respectively. We also note that for low enough angular momentum, oscillations at the inner shock 
influences that at the outer shock. For higher angular momentum, however, the perturbations are greatly washed out 
due to turbulence. As a result, only the inner shock oscillates and QPOs are produced at higher frequencies.

It is important to note that both the shocks are coupled together by 
the flow in between. As a result, both the shocks oscillate with higher
and lower frequencies. The behaviour of the power density spectra is very interesting.
In Fig. 10 we show the results of Case D where both the PDSs are 
superimposed. The solid curve shows the PDS of the time variation of the inner shock location
while the dotted curve shows the PDS of that of the outer shock.
It seems that the inner shock has more power at the higher frequency peak, while the 
outer shock has more power at the lower frequency peak. Assuming that the hard X-rays 
are produced due to Compton scattering of soft X-rays by hot electrons in the post-shock region,
this would mean that the hard X-rays will also modulate with high and low frequencies. The radiation 
from the  post-(outer)shock region would participate in the quasi-periodic oscillations
in the lower frequency and the radiation from the relatively small
 post-(inner)shock region would participate in the high frequency oscillation.

\begin{figure}
\includegraphics[height=12truecm,width=12truecm]{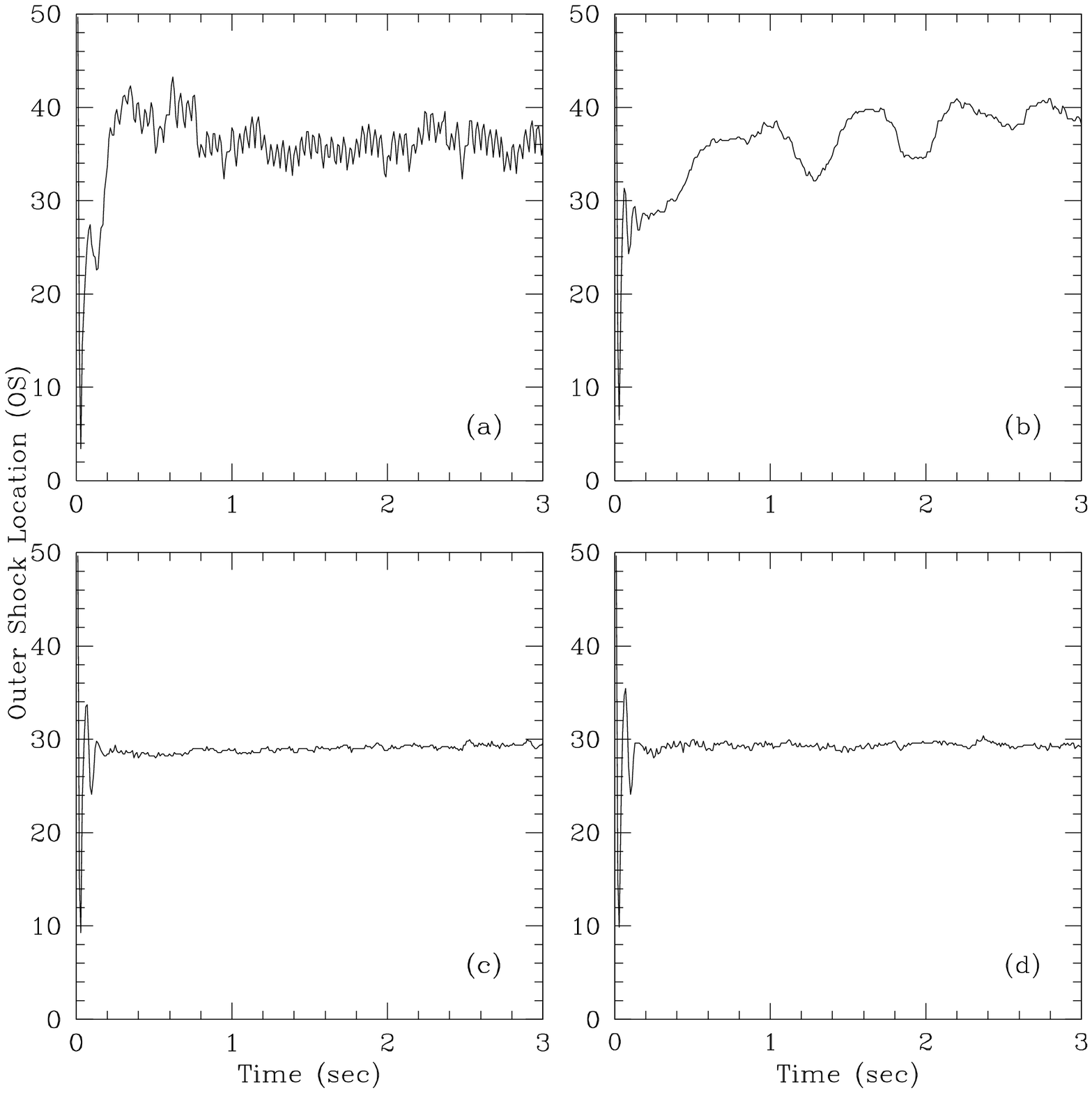}
\caption{{\bf(a-d):} Variation of the outer shock locations with time (seconds)
when the angular momentum is increased: (a) $\lambda=1.75$, (b) $1.85$, (c) $1.95$
and (d) $2.05$. The oscillation  frequencies are: (a) $\nu_{QPO}= 2.54$ Hz and (b) $2.93$, respectively.
There are no oscillations in (c-d) due to dominance of turbulence. }
\end{figure} 

\begin{figure}
\includegraphics[height=12truecm,width=12truecm]{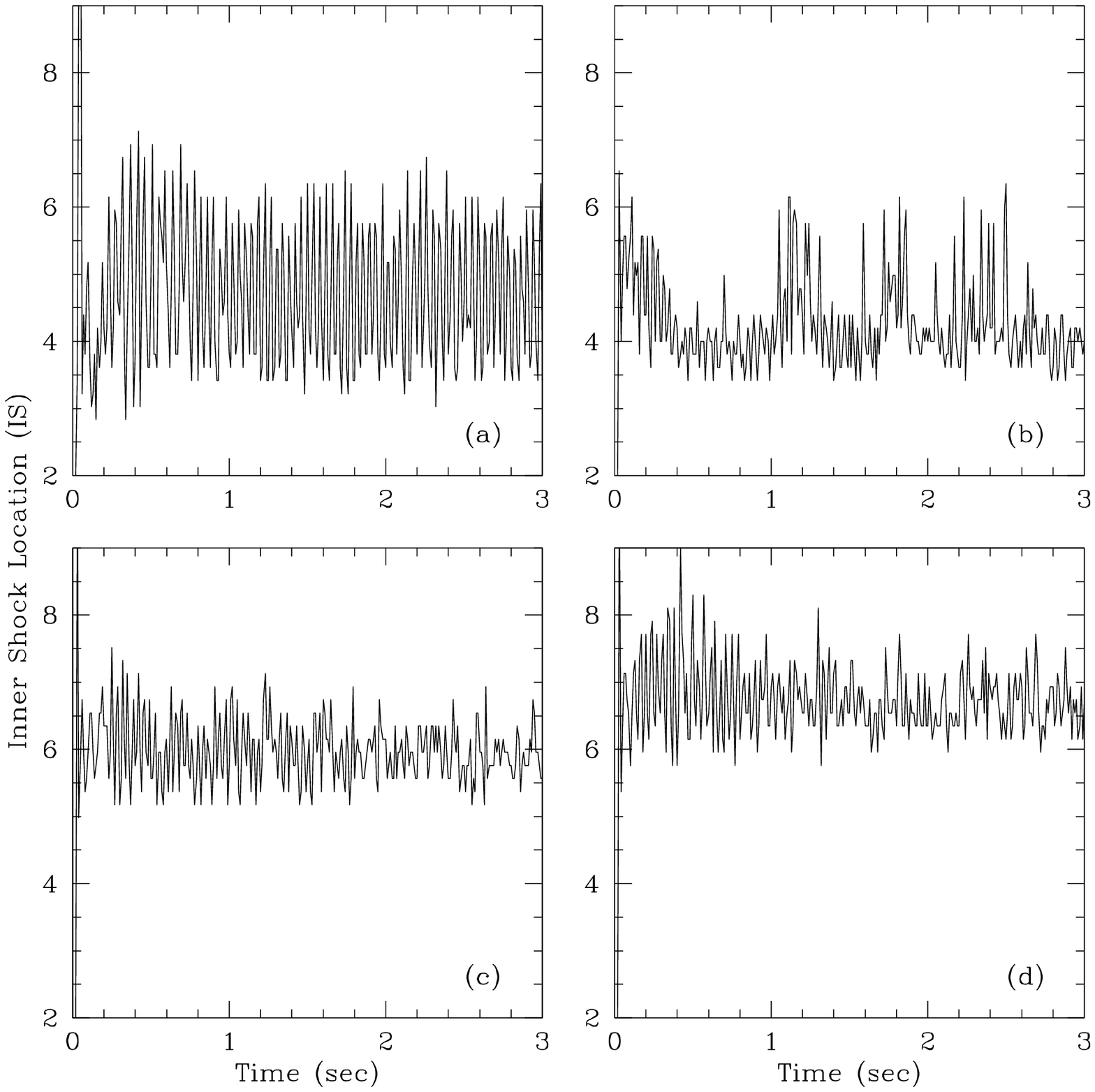}
\caption{{\bf (a-d):} Variation of the inner shock locations with time (seconds)
when the angular momentum is increased: (a) $\lambda=1.75$, (b) $1.85$, (c) $1.95$
and (d) $2.05$. The inner shock continues to oscillate in the entire range of angular momentum.
The oscillation frequencies are (a) $24.70$, (b) $25.50$, (c) $28.61$ and (d) $27.1$Hz respectively.}
\end{figure}

We turn our attention to the behaviour of the oscillating shocks where the specific 
energy of the flow is changed while the specific angular momentum is kept fixed.
The results  we show are those of  Cases (C-F) in Table 1.
In Figs. 11(a-d), we show the variation of the outer shock location
and in Figs. 12(a-d) we show the variation of the inner shock location. 
We assume $\lambda =1.75$ and ${\cal E}= 0.01,\ 0.02, \ 0.03, \ $ and $0.04$ 
were used. Since the shocks are primarily centrifugal barrier
supported, the locations of the shocks remain at similar distances,
though, we see a small increase in the mean shock (both the inner 
and the outer) locations with specific energy. This is expected from the 
theoretical point of view also. The frequency of oscillations of the 
outer (inner) shocks  are $ 2.4 (22.6), \ 2.54(24.7), \ 1.3(26.2)$ and $0.9, (23.8)$
respectively. The ratio $R_t$ of the infall time and free fall
(averaged over $20$ vertical grids above the equatorial plane) lies between $2$ to $3$.

\begin{figure}
\includegraphics[height=12truecm,width=12truecm]{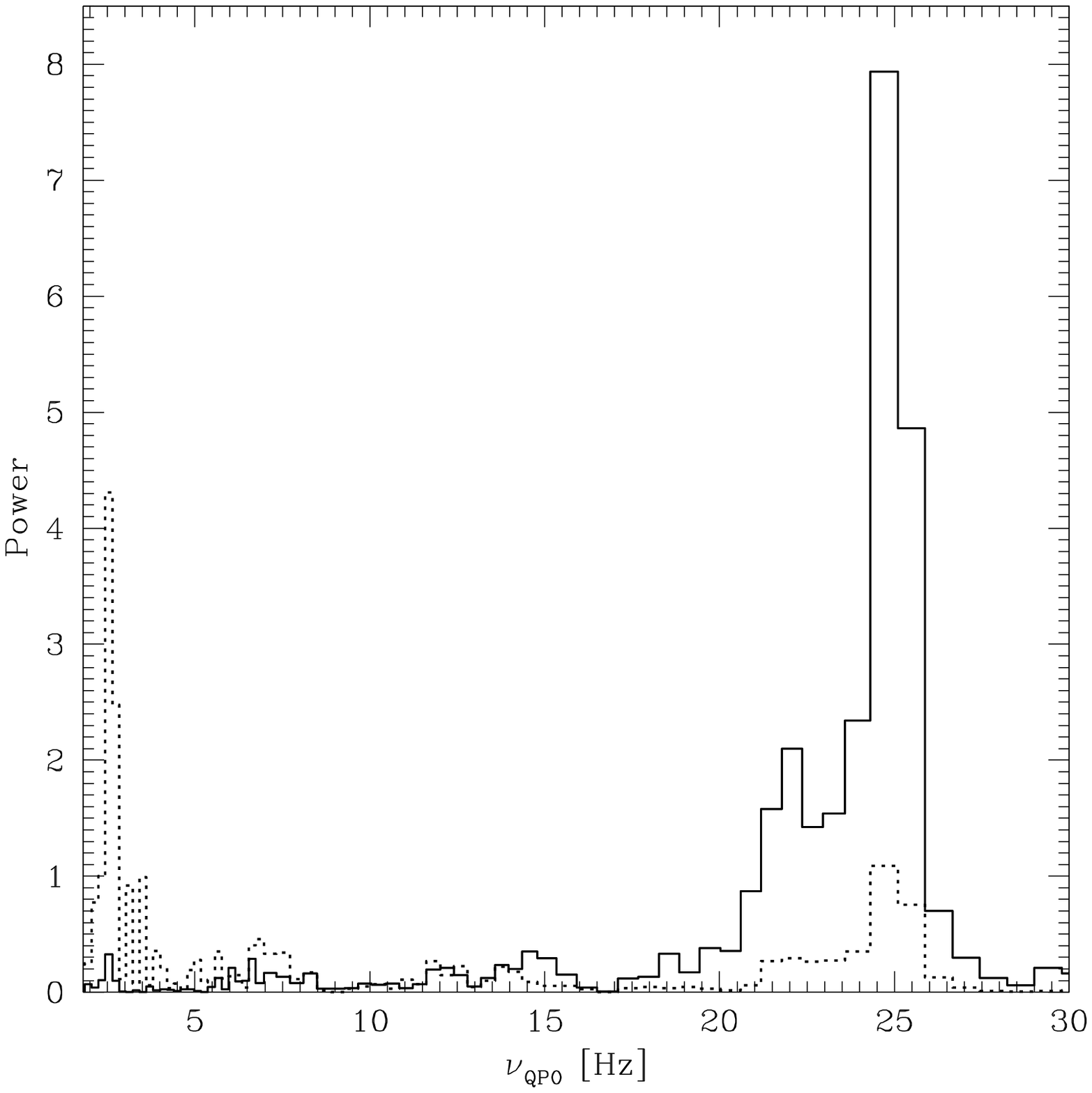}
\caption{Power density spectra of the time variation of the inner (outer) shocks
are shown by solid (dotted) histograms. Note that the PDS of inner shock has more power
at the higher frequency and that of the outer shock has more power at the lower frequency. } 
\end{figure} 

\begin{figure}
\includegraphics[height=12truecm,width=12truecm]{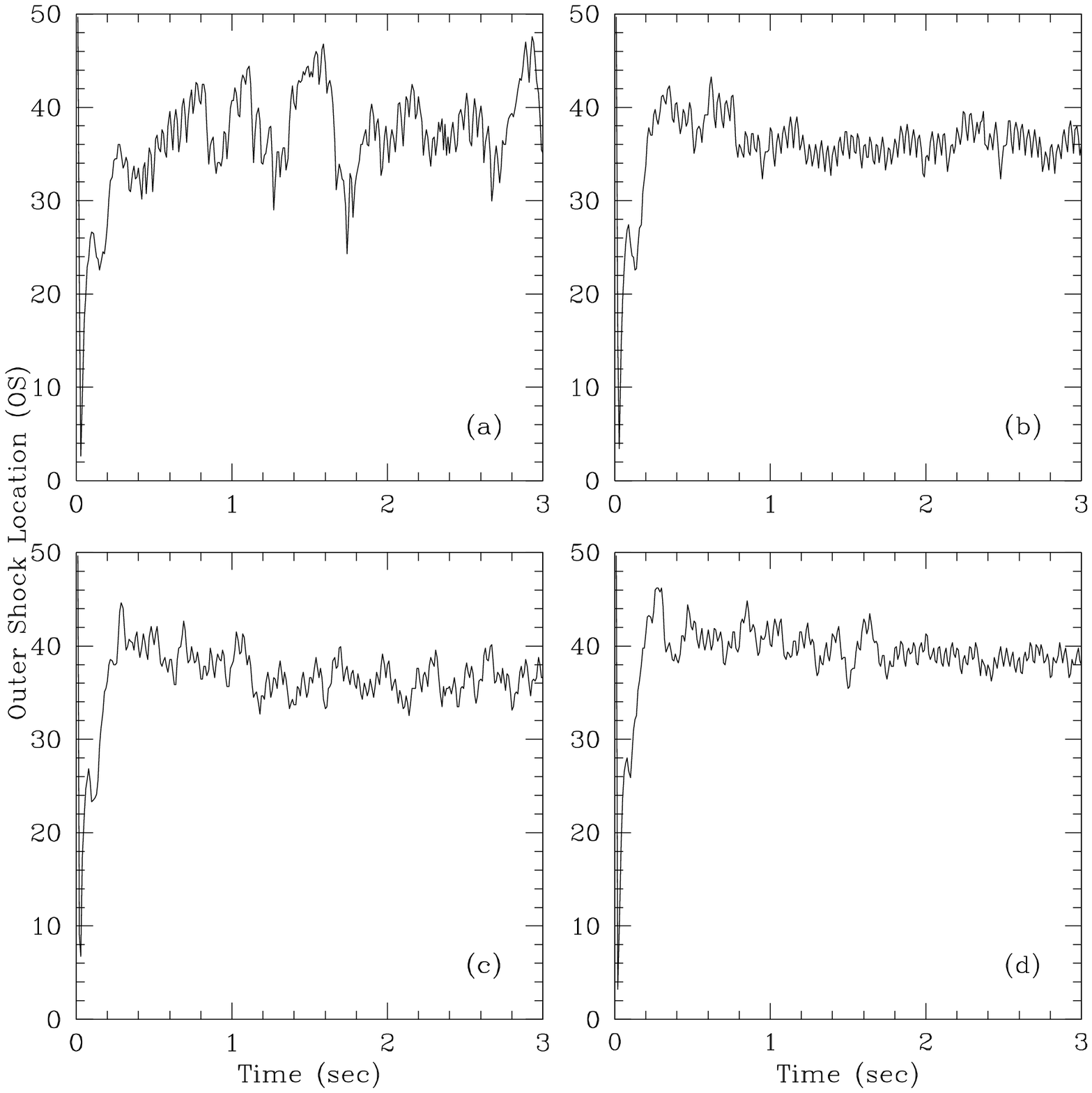}
\caption{{\bf(a-d):} Variation of the outer shock locations with time (seconds)
when the specific energy is increased: (a) ${\cal E}=0.01$, (b) $0.02$, (c) $0.03$
and (d) $0.04$. The specific angular momentum remains the same at $\lambda=1.75$.
The oscillation  frequencies are: (a) $\nu_{QPO}= 22.6$, (b) $24.7$, (c) $26.2$
and (d)$23.8$, respectively.} 
\end{figure} 

\begin{figure}
\includegraphics[height=12truecm,width=12truecm]{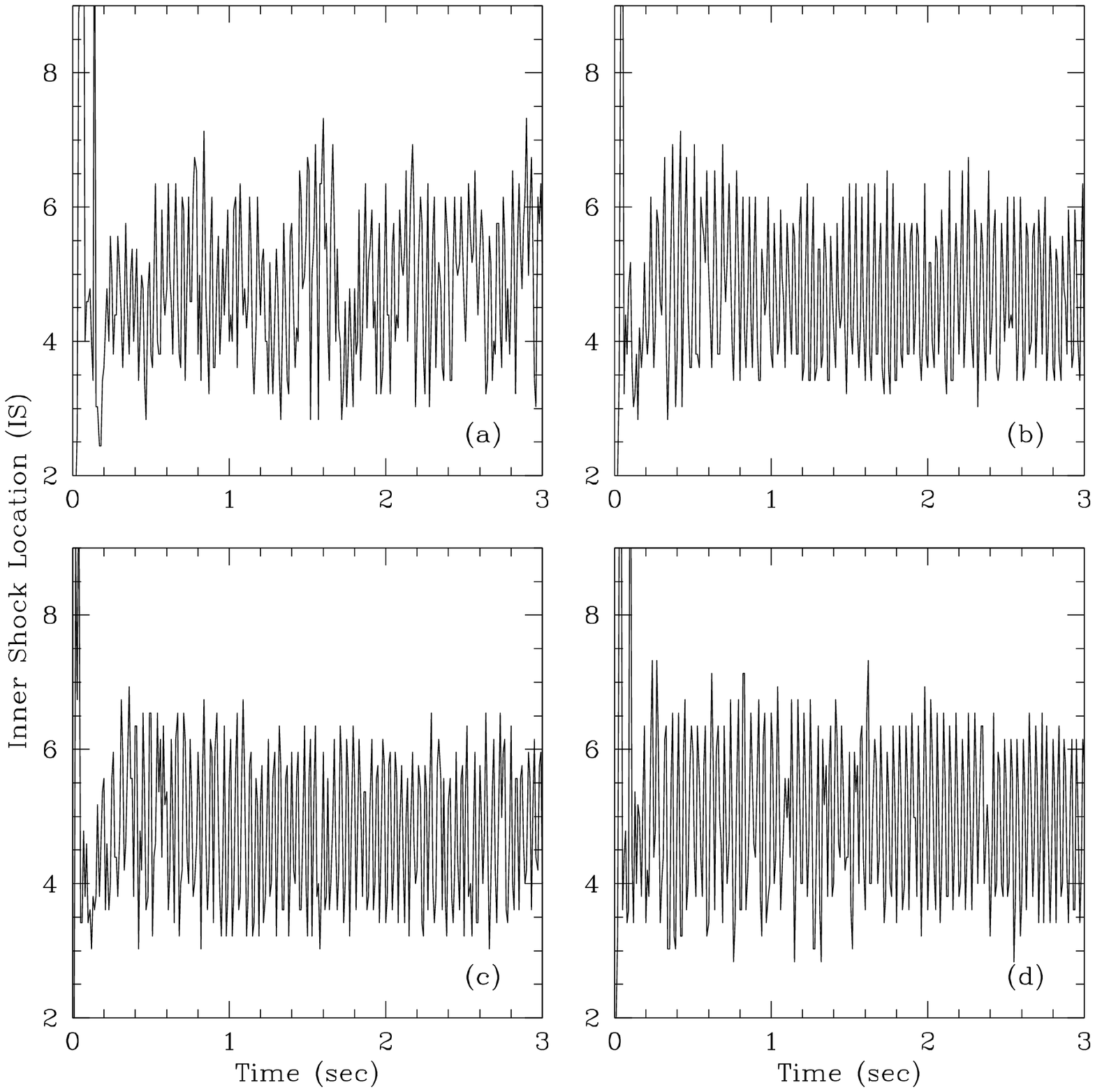}
\caption{{\bf(a-d)} Variation of the inner shock locations with time (seconds)
when the specific energy is increased: (a) ${\cal E}=0.01$, (b) $0.02$, (c) $0.03$
and (d) $0.04$. The specific angular momentum remains the same at $\lambda=1.75$.
The oscillation  frequencies are: (a) $\nu_{QPO}= 2.4$, (b) $2.54$, (c) $1.3$
and (d)$0.9$, respectively.}
\end{figure} 

We noted in Fig. 4 and 5(a-d) that a considerable amount of matter 
is ejected outwards after they are bounced from the centrifugal barrier.
It would be interesting to compute the amount of matter which leaves the grid
system due to outflows. In Figs. 13(a-b), we show the ratio of the outflow rate
(calculated by adding those matter leaving the grid) and inflow rate (calculated 
from the outer boundary condition) for the four Cases (Cases B, D, G, \& H) presented before. 
Here only $\lambda$ is varied. For clarity, we present the results for $\lambda =1.75$ and 
$1.85$ in Fig. 13a and those for $\lambda=1.95$ and $2.05$ in Fig. 13b. 
The mean ratios for the above Cases are $0.42$, $0.64$, $0.56$ and $0.58$ respectively.
We notice that the amplitude and frequency of the fluctuations in the outflow rate are 
mainly dictated by the fluctuations of the outer shock location, though the inner shock 
modulates it further.

\begin{figure}
\includegraphics[height=12truecm,width=12truecm]{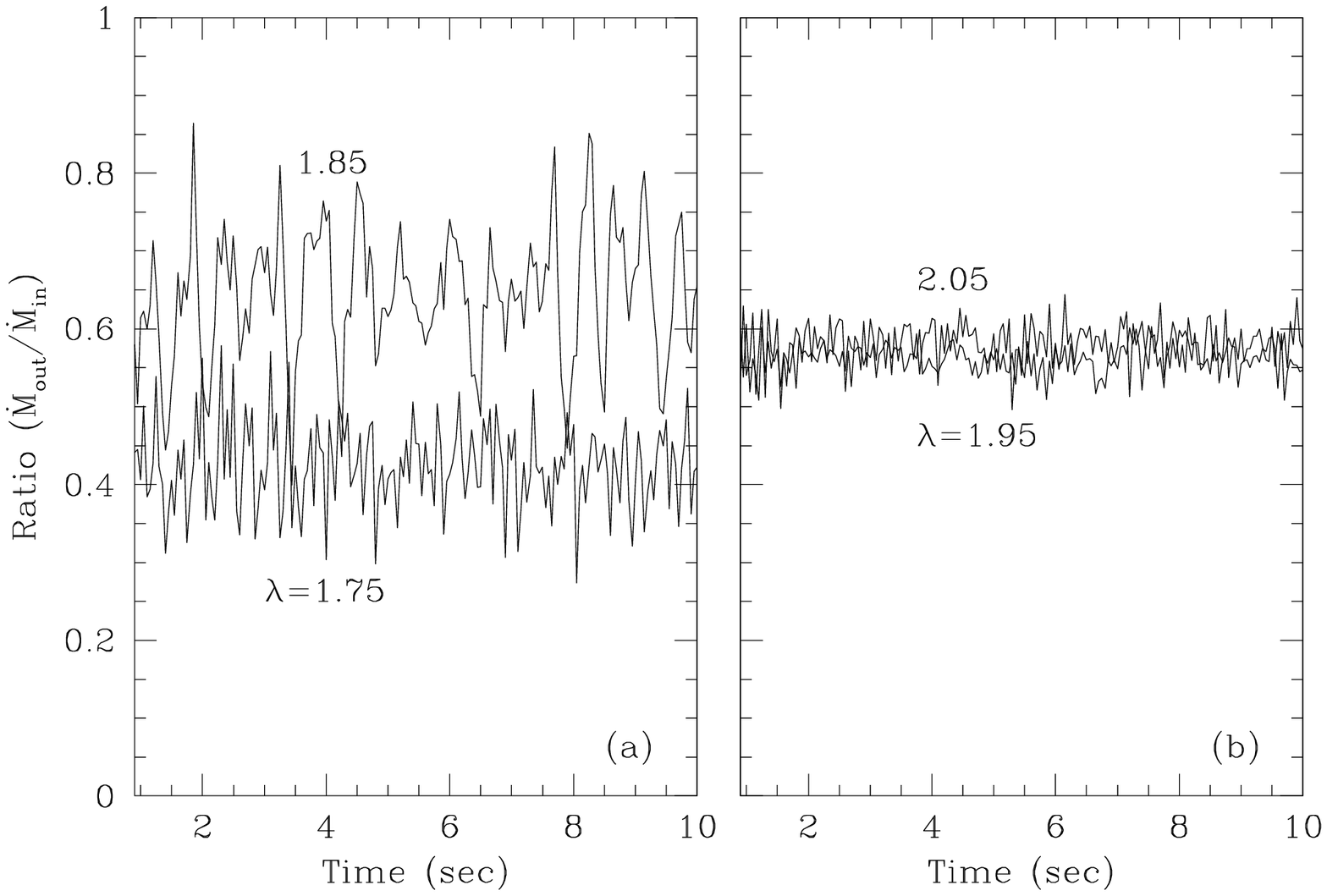}
\caption{Variation of ratio of outflowing matter to injected matter
with time when specific angular momentum is increased: (a) $\lambda=1.75, 1.85$,
 (b) $\lambda=1.95, 2.05$.}
\end{figure}

\section{Discussions and Concluding Remarks}

In this Paper, we presented the results of two dimensional hydrodynamic simulations
of matter accreting on to a black hole. We systematically chose the flow parameters 
(${\cal E}, \lambda$) from the parameter space which provides complete set of
solutions of a black hole accretion flow. The parameter space was classified into regions
which may or may not produce standing shocks (C89, C90) in an inviscid flow. The classifications
were made using three different models of the flow, namely, a disk of constant thickness,
a disk with conical wedge cross-section and a disk which is in vertical equilibrium
(Chakrabarti \& Das, 2001). Our motivation was to study whether a simulated result behaves like
a theoretical model throughout. We observed that the flow behaved
like that of constant thickness before the shock. However, in the post-shock region,
as the flow expands vertically due to higher thermal pressure, it behaved like 
that of a flow in vertical equilibrium. Second, the infall time in the post-shock region 
is several times larger as compared to the free-fall time, especially due to the formation
of turbulence in the post-shock region. Third, instead of only one possible
shock transition, the flow shows the formation of two shocks, one very close
to the black hole ($\sim 3-5$) and the other farther away depending on the angular 
momentum. Both the shocks showed significant oscillations.
While the inner shock oscillated faster than the outer shock, each of them 
also oscillated at the frequency of the other, though at a lesser power. These
oscillations or their variants are long thought to cause the quasi-periodic oscillations (QPOs)
observed in black hole intensity (Molteni, Sponholz \& Chakrabarti, 1996;
Chakrabarti, Acharyya \& Molteni, 2004). It is possible that not only the intermediate 
and low frequency QPOs are explained by this process, the high frequency QPOs 
may also be explained by the oscillations of the inner shock and the inner sonic point. However, since the 
volume of matter, participating in the inner shock oscillation is very small,
the modulation at a high frequency would be negligible. We also observed that the
outflows form from the post-shock region. The rates are specially high, and vary
episodically. The amplitude and frequency of variation of the outflow rate
is dictated by the amplitudes and frequencies of the two shocks. The outflow can be anywhere 
between $40$ to $80$ percent of the inflow rate, provided the flow is sub-Keplerian. 
Since the Keplerian flows are sub-sonic, and therefore, strictly speaking no shocks,
the outflows are not possible from a Keplerian disk in this model.

DR was supported by the Korea Research Foundation Grant funded by the Korean Government 
(MOEHRD) (KRF-2007-341-C00020). 

{}

\end{document}